\documentclass[11pt]{article}
\pdfoutput=1

\usepackage[]{Packages}
\bibliography{T-duality}

\usepackage{Definitions}

\newcommand*{\lattice}[3]{
    \begin{tikzpicture}[x=.5cm, y=.5cm]
    \coordinate (Origin)   at (0,0);
    \coordinate (XAxisMin) at (-1,0);
    \coordinate (XAxisMax) at (3,0);
    \coordinate (YAxisMin) at (0,-1);
    \coordinate (YAxisMax) at (0,3);
    \clip (-1,-1) rectangle (5,5); % Clips the picture...

    \draw[style=help lines,dashed] (-2,-2) grid[step=1] (7,7);

    \pgftransformcm{#1}{0}{#2}{#3}{\pgfpoint{0}{0}}

    \coordinate (Xone) at (1,0);
    \coordinate (Xtwo) at (0,1);

    \foreach \x in {-7,-6,...,7}{% Two indices running over each,
      \foreach \y in {-7,-6,...,7}{% node on the grid we have drawn 
        \node[draw,circle,inner sep=1.5pt,fill] at (1*\x,1*\y) {};
      }
    }
    \filldraw[fill=gray, fill opacity=0.3, draw=black] (Origin)
        rectangle ($(Xone)+(Xtwo)$);

    \draw[style=help lines,dashed, red] (-2,-2) grid[step=1] (7,7);

    \end{tikzpicture}
}

\preprint{}

\newcommand{\OfficialTitle}{Phases of N=2 Necklace Quivers
}
\title{\vspace{2cm}
  {\color{Thoughtless}\Huge\textbf{\dosserif\OfficialTitle}}
}

\hypersetup{pdfauthor={Antonio Amariti and Domenico Orlando and Susanne Reffert},pdftitle={\OfficialTitle}}

\author{%
  \begin{minipage}{.8\linewidth}
    \vspace{1cm}
    \begin{center} \dosserif
      {\small 
        \textbf{Antonio~Amariti}\textsuperscript{\ding{95}},
         \textbf{Domenico~Orlando}\textsuperscript{\ding{96}} and 
        \textbf{Susanne~Reffert}\textsuperscript{\ding{96}}}
    \end{center}
    \vspace{1cm}
     \authorBlock{\ding{95}}{ Physics Department, The City College of the \textsc{cuny},\\
160 Convent Avenue, New York, \textsc{ny} 10031, \textsc{usa}}
    \authorBlock{\ding{96}}{Albert Einstein Center for Fundamental Physics\\
      Institute for Theoretical Physics\\
      University of Bern,\\
      Sidlerstrasse 5, \textsc{ch}-3012 Bern, Switzerland}
  \end{minipage}
}

\date{} 

\begin{document}

\setstretch{1.15}

\numberwithin{equation}{section}

\begin{titlepage}

  \newgeometry{top=23.1mm,bottom=46.1mm,left=34.6mm,right=34.6mm}

  \maketitle

  \thispagestyle{empty}

  \vfill\dosserif
  
  \abstract{\normalfont \noindent We classify the phases of 
    $\mathcal{N}=2$ elliptic models in terms of their global
    properties \emph{i.e.} the spectrum of line operators. We show the
    agreement between the field theory and the M--theory analysis and
     how the phases form orbits under the action of the
    S--duality group which corresponds to the mapping class group of the
    Riemann surface in M--theory. }

\vfill

\end{titlepage}

\restoregeometry

%%%%%%%%%%%%%%%%%%%%%%%%%%%%%
% \tableofcontents

%\listoffigures

% \listoftodos

%%%%%%%%%%%%%%%%%%%%
\section{Introduction}
\label{intro}

In this note, we study the charge lattices of mutually local bound states of
Wilson and 't~Hooft lines for $\mathcal{N}=2$ elliptic models, corresponding to chains of $A_{N-1}$
gauge groups connected by bifundamental hypermultiplets~\cite{Witten:1997sc}.
% These lattices provide the additional data necessary to fix the
% structure of the gauge group.
We first study the problem in a
field theory description by considering the models in the $\mathcal{N}=1$ formalism.
% In the second part of the paper 
Then we reproduce the results in M--theory, %.
where the models are obtained by 
% In this geometric picture the models correspond to 
wrapping an \M5--brane \(N\) times on a punctured torus.
%In this geometry,
% T
The charges of the line operators become homologies of closed curves % in such geometry.
and the lattices are reproduced % The analysis of the covering geometry reproduces the lattice through
in terms of 
the fundamental group of the 
surface.
The geometric description is useful for understanding the action of the
S--duality group on the lattices in terms of the mapping class group of the 
punctured torus.

 \medskip
The phases of  $\mathcal{N}=4$  \ac{sym} can be classified in terms of the 't~Hooft classification of the possible 
vacua~\cite{'tHooft:1977hy,'tHooft:1979uj,'tHooft:1981ht}.
The analysis can be further extended to the cases with $\mathcal{N}<4$  by adding a
supersymmetry-breaking mass deformation~\cite{Donagi:1995cf} . 
The classification boils down to determining the  maximal
charge lattice of mutually local bound states of electric \acp{Wline}  and magnetic \acp{Hline} (see \cite{Kapustin:2005py} for a precise definition
of these operators).
The charges are taken with respect to the center of the gauge group and the mutual locality
constraints correspond to a generalized \ac{dsz} quantization condition.
In recent years, this subject has returned to the spotlight of interest due to the discovery of the relation between these lattices and the global properties of the gauge group~\cite{Gaiotto:2010be,Aharony:2013hda}.
The gauge group of a quantum field theory is fixed when the
gauge algebra is supplemented with additional data such as the charge lattices discussed above.

In the four-dimensional \(A_{N-1}\) $\mathcal{N}=4$ \ac{sym} theory, each lattice corresponds to a 
phase of the $SL(2,\setZ)$ S--duality group, thus realizing  a representation that is in general reducible. 
In other words, the lattices can be organized in (disjoint) orbits under S--duality.
% ❌ The lattices are usually related to each other by the generators 
% ❌ $S$ and $T$ of  $SL(2,\setZ)$. There are also cases in which there are orbits
% ❌ in the $SL(2,\setZ)$ duality group, \emph{i.e.} where it is not possible to connect all the lattices
% ❌ by the action of $SL(2,\setZ)$.
This problem has been reformulated in M--theory in~\cite{Amariti:2015dxa}: in this language, the gauge theory lives on 
\M5--branes wrapping the M--theory torus $N$ times, and the bound states are \M2--lines wrapping the covering geometry.
The problem of computing the possible lattices on the field theory side is translated into the study of the intersections
of the closed \M2--lines.
Indeed, by associating the homologies of these curves to the charges of the lines in field theory, 
one obtains the \ac{dsz} quantization condition and recovers the expected charge spectrum.

\bigskip

A similar situation is expected in four-dimensional $\mathcal{N}=2$ gauge theories arising from wrapping \M5--branes on
Riemann surfaces. 
So far, only the case of non-Lagrangian class S theories~\cite{Gaiotto:2009we} 
has been discussed in the literature~\cite{Drukker:2009tz,Tachikawa:2013hya,Xie:2013vfa,Amariti:2016bxv}.
These theories can be regarded as the low-energy description of the dynamics of $N$ \M5--branes compactified on genus $g$ Riemann surfaces with $r$ punctures,
$\Sigma_{g,r}$. 
The case of $r=0$ has been reformulated in~\cite{Amariti:2016bxv} in terms of the homologies 
of closed lines on the Riemann surface, while case with punctures has not been fully explored yet.
A  systematic analysis of the punctured case can however be initiated on a
simpler, Lagrangian class of  $\mathcal{N}=2$ gauge theories.
It corresponds to the so-called elliptic models of~\cite{Witten:1997sc}, $\mathcal{N}=2$ Lagrangian gauge theories
with product gauge group on a necklace quiver.
%
% Coming back to the discussion above, i
It is natural to expect that this generalization will lead to a classification of
the phases similar to the one discussed in $\mathcal{N}=4$ \ac{sym}.
This intuition comes from the fact that 
% The reason beyond this claim is that 
the case with one puncture corresponds to the 
$\mathcal{N}=2^*$ theory studied in~\cite{Donagi:1995cf}, where it was observed that all the  
 phases present in $\mathcal{N}=4$ persist after the mass deformation is switched on.

\bigskip

Motivated by this analogy, in this paper we study the phases   of the $\mathcal{N}=2$ elliptic 
models. 
In the first part of our analysis, in section \refstring{field}, we study the problem
in a purely $\mathcal{N}=1$ field-theoretical approach.
We consider a general quiver with $r$ nodes and compute the charge lattices of the bound states of 
\ac{WH} lines by imposing a generalized \ac{dsz} condition. The presence of bifundamental
hypermultiplets connecting the nodes of the quiver imposes additional constraints on the allowed  $2 r $-dimensional lattices. 
We show that the possible lattices are actually two dimensional and -- as expected -- coincide with the
ones obtained in $\mathcal{N}=4$ \ac{sym}.
The second part of the analysis,  presented in section \refstring{geometry}, focuses on the M--theory description . In this picture, we have a genus one Riemann surface with $r$ 
punctures, $\Sigma_{1,r}$. We show that the analysis of the homologies of closed \M2--lines in this geometry reproduces the field theory results.
As already observed in~\cite{Amariti:2015dxa}, also in this case the quantum constraint imposed on the field theory side (the \ac{dsz} condition) is a classical phenomenon in the geometric description.

The M--theory analysis has the advantage of giving a simple realization for the action of the S--duality group,
corresponding to the mapping class group of the punctured Riemann surface \(\Mod(\Sigma_{1,g})\)~\cite{Witten:1997sc} (see also~\cite{Hanany:1998ru,Halmagyi:2004ju} for related discussions).
In section  \refstring{duality}
we study the action of this group on the geometric side and translate its action on the charges
of the bound states of line operators. 
The net effect is that a part of the S--duality group, generating an $SL(2,\setZ)$ subgroup% of the mapping class group
,
acts on the lattices as in the case of $\mathcal{N}=4$ \ac{sym}, while the rest of the action
leaves the lattices invariant.

An explicit example, namely the one of the quiver \(A_1 \oplus A_1 \oplus A_1 \) is discussed in section \refstring{sec:ex} and further directions are discussed in section \refstring{conclusions}.

%%%%%%%%%%%%%%%%%%%%%
\section{Global properties of elliptic models}
\label{field}

\begin{figure}
  \centering
  \begin{tikzpicture}[]
    \node at (0,0) {  \includegraphics[width=5cm]{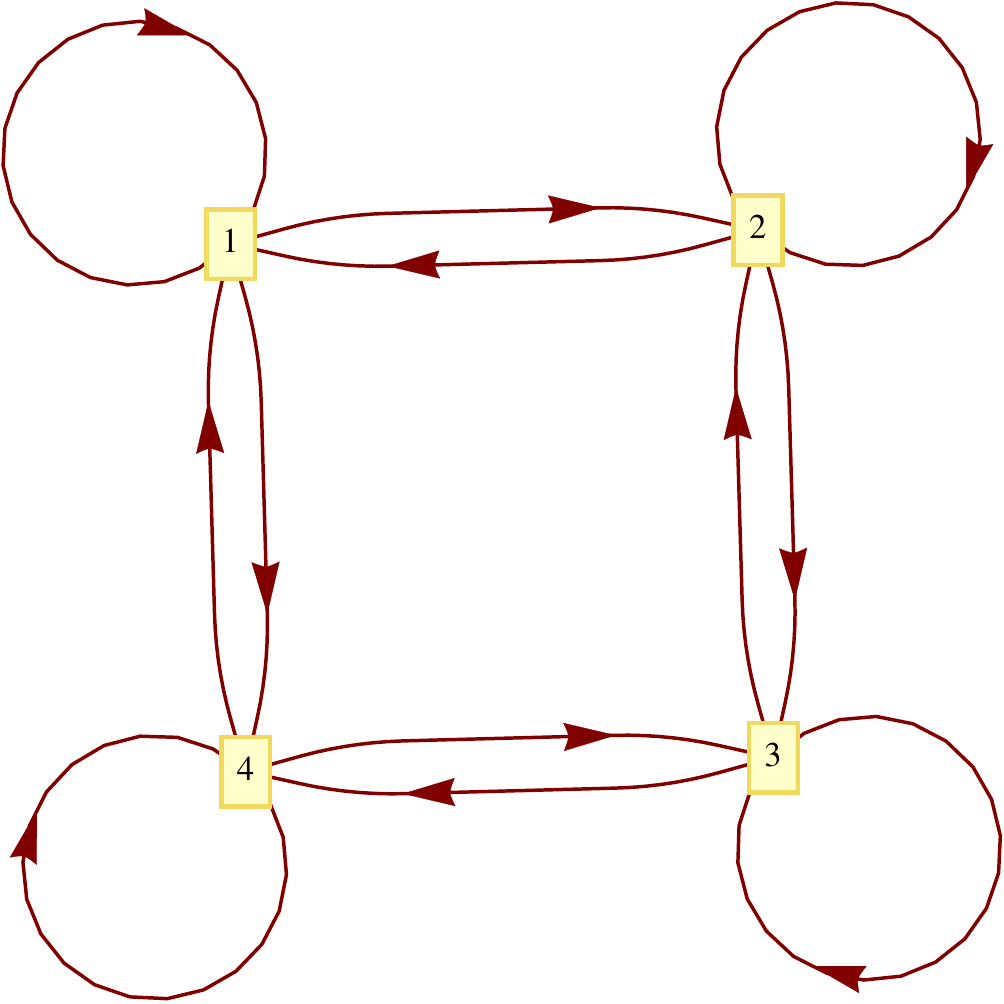}};
    \draw[fill=yellow!60] (-1.6,1.6) rectangle (-1,1) node[pos=.5] {\(1\)};
    \draw[fill=yellow!60] (1.6,1.6) rectangle (1,1) node[pos=.5] {\(2\)};
    \draw[fill=yellow!60] (1.6,-1.6) rectangle (1,-1) node[pos=.5] {\(3\)};
    \draw[fill=yellow!60] (-1.6,-1.6) rectangle (-1,-1) node[pos=.5] {\(4\)};
  \end{tikzpicture}

  \caption{Quiver description of an elliptic model with $r=4$ in
    $\mathcal{N}=1$ notation. Each node represents an $\mathcal{N}=2$
    vector multiplet.%, obtained by combining an $\mathcal{N}=1$ vector multiplet and an $\mathcal{N}=1$ multiplet in the adjoint representation.
    Each pair of arrows connecting a pair of consecutive nodes
    represents a bifundamental $\mathcal{N}=2$ hypermultiplet.
    %,represented in the quiver as a pair of $\mathcal{N}=1$
    %bifundamental chiral multiplets, two arrows with the opposite
    %orientation.
    }
    \label{quiver}
\end{figure}
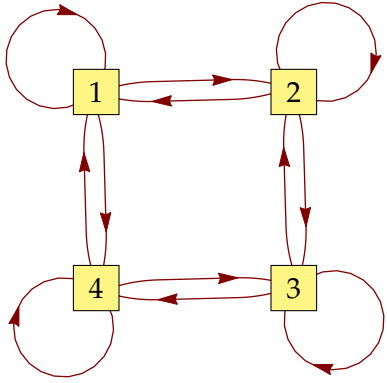

In this section, we study the global properties of an infinite class of
$\mathcal{N}=2$ gauge theories with $r$ gauge groups.
These gauge theories can be represented conveniently via a quiver diagram.
One can associate each gauge group to a node and place the nodes on a circle.
Each pair of consecutive nodes is connected by two arrows with opposite orientations.
These arrows represent a pair of bifundamental $\mathcal{N}=1$ chiral fields
$X_{l,l+1}$ and $X_{l+1,l}$, \emph{i.e.} the $\mathcal{N}=2$ hypermultiplets.
There is also an $\mathcal{N}=1$ adjoint field $X_{l,l}$ associated to each node, corresponding to the $\mathcal{N}=2$ vector multiplets.
In Figure~\ref{quiver}, an example of such a quiver
 with $r=4$ is shown.
 The matter fields interact through a superpotential 
\begin{equation}
W = \sqrt{2} \sum_{l=1}^r (X_{l,l+1} X_{l+1,l+1} X_{l+1,l} - X_{l+1,l} X_{l,l} X_{l,l+1})
\end{equation}
where the sum is cyclic (the label $l= r+1$ is identified with $l=1$) and the coupling is fixed by supersymmetry.
We consider the case in which each gauge component has algebra $A_{N-1}$ and the full gauge group has the form
\begin{equation}
  G = \frac{\prod_{l=1}^{r} SU(N)_l \times U(1)}{\setZ_N} ,
\end{equation}
where $\mathbb{Z}_N$ is diagonally embedded
\cite{Aharony:1998qu,Witten:1998wy,Belov:2004ht}.
In the \ac{ir}, the overall $U(1)$ gauge symmetry decouples from the dynamics.
The different consistent factorizations of this $U(1)$ symmetry correspond to the 
different possible choices of the global properties of the gauge
group~\cite{Moore:2014gua}\footnote{We would like to thank Ofer Aharony for pointing out
  this fact to us.}. 
  %We find that this group has the form
%\begin{equation}
   % G_{k,i} \equiv  \left( \frac{\prod_{l=1}^{r} SU(N)_l}{\setZ_k}\right)_i,
%\end{equation}
%as we explain in the following.

\bigskip

% \medskip
% In the rest of this section, we
We can discuss these different possibilities 
by studying the charge lattice of the mutually local bound states
of the \ac{WH} lines.
A \ac{Wline} \(W_l\) and a \ac{Hline} \(H_l\) can be introduced for each $(A_{N-1})_l$ gauge component% ❌ , 
% denoted by H$_j$ and W$_j$
. Let \(e_l\) be the charge of the \ac{Wline}
under the center $\setZ_N$ 
 of the $l$-th $A_{N-1}$  factor 
and  $m_l$ be the charge of the related \ac{Hline}. 
We refer to the charge $e_l$ as \emph{electric} charge and to the charge $m_l$
as \emph{magnetic}.
A generic line operator in this quiver corresponds to a combination of 
$W_l$ and $H_l$ lines. We denote such an operator as $(W_1,\dots,W_r;H_1,\dots,H_r)$ and
 its charge vector is
\begin{equation}
 l_\mathcal{O} =(e_1,\dots,e_r; m_1,\dots,m_r).
\end{equation}
These charges define a $\pqty{\setZ_N}^r \times \pqty{\setZ_N}^r$ lattice and each point of this lattice is associated to a class of $(W_1,\dots,W_r;H_1,\dots,H_r)$
bound states. Each pair of such states has to be mutually local. 
This  is equivalent to imposing a \ac{dsz} condition on the lines.
For a pair of lines $( e_1, \dots, e_r; m_1, \dots, m_r )$ and  $( e_1',  \dots, e_r'; m_1', \dots, m_r' )$,
the condition is 
\begin{equation}
  \label{DSZ-general-An}
  \sum_{l=1}^{r} e_l m_l' - e_l' m_l = 0 \mod N.
\end{equation}

In $\mathcal{N}=4$ \ac{sym}, the spectra of line operators are determined by imposing 
this condition on the charges.
Here this is not enough: the conditions must be supplemented by some information on the structure of the quiver because
the  bifundamental matter is not compatible with some of the lattices that solve the \ac{dsz} quantization.
In order to construct the lattices, we can set up the problem as follows.
Consider a bifundamental field  $X_{l,l+1}$ charged under the \(l\)-th and the \(l+1\)-st group: this corresponds to a line operator
where $e_l=-e_{l+1}=1$ and all other charges are set to zero. Imposing the \ac{dsz} condition between \(X_{l,l+1}\) and a generic line we find
\begin{equation}
  m_l - m_{l+1} = 0 \mod N.
\end{equation}
Applying this constraint to the rest of the quiver, we find $m_l = m \mod  N$ for each value of $l$.
This is the first simplification and we can now express the charge of a line operator as
\begin{equation}
 l_\mathcal{O} =  (e_1, \dots, e_r; m, m, \dots) = (e_1,\dots,e_r; m) .
\end{equation}
The \ac{dsz} condition in Eq.~\eqref{DSZ-general-An} becomes
\begin{equation}
  \label{DSZ-with-hypers}
  \pqty{\sum_{l=1}^{r} e_l} m' - \pqty{\sum_{l=1}^{r} e_l'} m = 0 \mod N.
\end{equation}

A second simplification is possible because 
by linearity, the existence of two lines with charges $e_1$ and $e_2$ implies the existence of a
line with charge $e_1' = e_1+e_2$, \(e_2' = 0\).
Let \(l\) be the line $l=(e_1, e_2, 0, \dots ; 0)$. In the theory, there is always the line $l_{X_{1,2}} = (1,-1,0, \dots ; 0)$  that has the same charge as the bifundamental  field
$X_{1,2}$. This means that by linearity, the charge $l + e_2 l_{X_{1,2}} = (e_1 + e_2, 0, \dots; 0)$ is also allowed.
In general, if there is a line $(e_1,\dots,e_r;m)$, there is also a line $(\sum_r e_r,0,\dots,0;m)$ and we can use this line as a representative for the whole family.
We conclude that a generic line belongs to a family parametrized by a pair of
integer charges, $ l_\mathcal{O} =(e;m)$ where \(e\) is the sum of the electric charges and \(m \) is the unique magnetic charge. The \ac{dsz} condition in Eq.~\eqref{DSZ-with-hypers} becomes a condition 
on the charges $(e;m)$ and $(e',m')$, \emph{viz.}
\begin{equation}
  e m'-m e' =0 \mod N.
\end{equation}
We have just reformulated the lattice $\pqty{\setZ_N}^r \times \pqty{\setZ_N}^r$
as a $\setZ_N \times \setZ_N$ lattice.
A two-dimensional lattice is generated by two non-negative integer vectors $(k,0)$ and $(i,k')$, where $k k' = N$
and $0 \le i < k$.
Once these two integers are specified, the global gauge group is fixed.
We denote the gauge group by
\begin{equation}
  G_{k,i} \equiv  \left( \frac{\prod_{l=1}^{r} SU(N)_l}{\setZ_k}\right)_i,
\end{equation}
where the choice of $k$ fixes the quotient $\setZ_k$
and the integer $i$ is the electric charge of the line with the lowest possible non-vanishing magnetic charge $m = N/ k$. This shows that the lattice structure of the $\mathcal{N}=2$ elliptic models
is identical to the one of $\mathcal{N}=4$ \ac{sym}.

For $\mathcal{N}=4$ \ac{sym}, the different possible lattices for a given algebra $A_{N-1}$ can be arranged into representations of the $SL(2,\setZ)$
symmetry acting on the gauge coupling. 
%This symmetry corresponds to the S--duality group, an exact
%symmetry of the \tIIB string theory. 
%It acts on the holomorphic gauge coupling $\tau$ by the combined effect of the generators $T$ and $S$ 
%of $SL(2,Z)$,  $S: \tau \rightarrow -1/\tau$ and $T:\tau \rightarrow \tau +1$. \todos{kill?}
%In the case of the $\mathcal{N}=2$ elliptic models, the situation is more complicated.
%Each gauge component has a holomorphic coupling $\tau_l$ but the coupling of the \tIIB string theory 
%is the sum $\tau = \sum_{l=1}^{r} \tau_l$.
%One may be tempted to associate the $SL(2,\setZ)$ symmetry group to this last coupling, but this is too naive.
%The situation can be better visualized in $M$--theory.
%
%\medskip
In the next section, we will derive the lattices from the $M$--theory description and study the action of the S--duality
on the geometry.
After that, we will translate this action into the field theory language and study its effect on the charge lattices.

%%%%%%%%%%%%%%%%%
\section{Geometry}
\label{geometry}

In this section we rederive the field theory results obtained in the last section via M--theory.
The M--theory description of the elliptic models has been originally discussed in~\cite{Witten:1997sc}
as an uplift of the \tIIA description. The latter consists of 
a stack of \(N\) \D4--branes extended along $x^{0123}$ and wrapping the compact direction $x^6$.
There are also $r$ parallel \NS5--branes, extended along $x^{012345}$,  
placed at the positions $p_l = x^6_l$.

The lift to M--theory happens along the coordinate $x^{10}$. The \(N\) \D4--branes
branes by themselves would become an \M5--brane wrapping $N$ times the two compact directions $x^6$ and $x^{10}$, while
the \NS{} branes lift to \M5--branes at fixed positions in $x^6$ and $x^{10}$. 
Together, the geometric picture consists of the \(N\)-cover of $\Sigma_{1,r}$, a genus one Riemann  
surface with \(r\) punctures. We refer to this covering geometry as  $\Sigma^N_{1,r}$.
%   i.e. they can be considered as point-like objects in this 
% description and they can be represented as marked points or punctures on the M--theory torus.
By ordering the punctures, one can interpret the distance between two consecutive punctures
along $x^{6}$ and $x^{10}$ as the holomorphic gauge coupling of a node of the quiver
of the four-dimensional theory:
\begin{equation}
  \begin{aligned}
    \tau_l &=  \frac{i(x_{l+1}^{6} - x_{l}^6)}{16 \pi^2 g_s L} + \frac{x_{l+1}^{10} - x_{l}^{10}}{2 \pi R}, & l =1,\dots,r-1 \\
    \tau_r &=  \frac{i(x_{1}^{6} - x_{r}^6 + 2 \pi L)}{16 \pi^2 g_s L} + \frac{x_{1}^{10} - x_{r}^{10}+\theta R}{2 \pi R}, 
  \end{aligned}
\end{equation}
% \begin{eqnarray}
% \tau_i &=&  \frac{i(x_{i+1}^{6} - x_{i}^6)}{16 \pi^2 g_s L} + \frac{x_{i+1}^{10} - x_{i}^{10}}{2 \pi R} 
% \quad \quad i=1,\dots,r-1 \\
% \nonumber \\
% \tau_r &=&
%  \frac{i(x_{1}^{6} - x_{r}^6 + 2 \pi L)}{16 \pi^2 g_s L} + \frac{x_{1}^{10} - x_{r}^{10}+\theta R}{2 \pi R} 
% \end{eqnarray}
where the periodicity in the coordinates $x_6$ and $x_{10}$ is respectively
$2 \pi L$ and $\theta R$. 

In the previous section we have studied % On the field theory side, 
the global properties by supplementing the theory with additional data, the charges of the line operators.
A \ac{Wline} or a \ac{Hline} is represented in the geometric picture by an \M2--brane extended in \(x^0\) (the time direction), \(x^4\) (a direction perpendicular to the \M5--brane) and wrapping a geodesic on the Riemann surface \(\Sigma^N_{1,r}\).  Such \M2--branes appear as lines on $\Sigma^N_{1,r}$ and we refer to them as \M2--lines.
An \M2--line extended in \(x^{10}\) and at fixed \(x^6\) passing between two punctures \(P_l\) and \(P_{l+1}\) corresponds to a \ac{bps} state with electric charge \(e_l = 1\), while any  \M2--line extended in \(x^6\) and at fixed \(x^{10}\) is a state with magnetic charge \(m = 1\).
More in general, the charges of the line operators on the field theory side correspond to  the homologies of the closed  \M2--lines on \(\Sigma_{1,r}\).

\begin{figure}
  \centering
  \begin{tikzpicture}
    \node at (0,0) {\includegraphics[width=.5\textwidth]{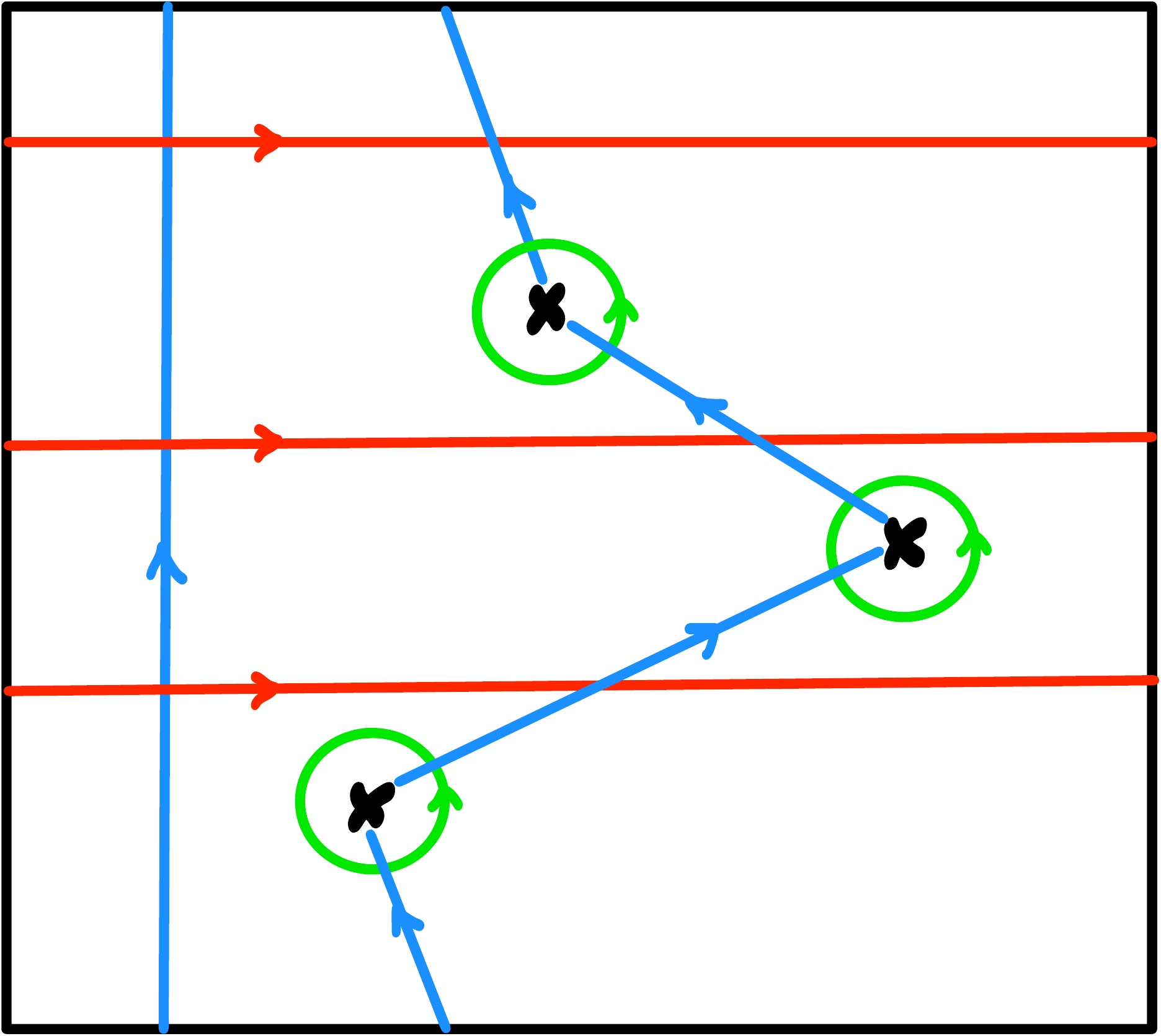}};
    \node at (-2,2.6) {\(\alpha \)};
    \node at (-2,.8) {\(\alpha_1\)};
    \node at (-2,-.7) {\(\alpha_2\)};

    \node at (-2.8, -.2) {\(\beta\)};
    \node at (0,2) {\(\beta_3\)};
    \node at (-.7,-2.3) {\(\beta_3\)};
    \node at (1,1) {\(\beta_1\)};
    \node at (.8,-.3) {\(\beta_2\)};

    \node at (.6,1.3) {\(\gamma_1\)};
    \node at (2.7,-.1) {\(\gamma_2\)};
    \node at (-.5,-1.7) {\(\gamma_3\)};

    % \node at (-.6,1.3) {\(P_1\)};
    % \node at (2.4,-.1) {\(P_2\)};
    % \node at (-1.7,-1.7) {\(P_3\)};

    % \node at (.2,-.7) {\(\delta_2\)};
    % \node at (-1.2,-.7) {\(\delta_1\)};

    % \node at (-1.2,-2.5) {\(\gamma_3\)};
    % \node at (2.8,0) {\(\gamma_2\)};
    % \node at (0.5,1.5) {\(\gamma_1\)};

  \end{tikzpicture}
  \caption{Cycles and paths used in this note for the torus with \(r = 3\) punctures.}
  \label{fig:cycles-and-paths}
\end{figure}
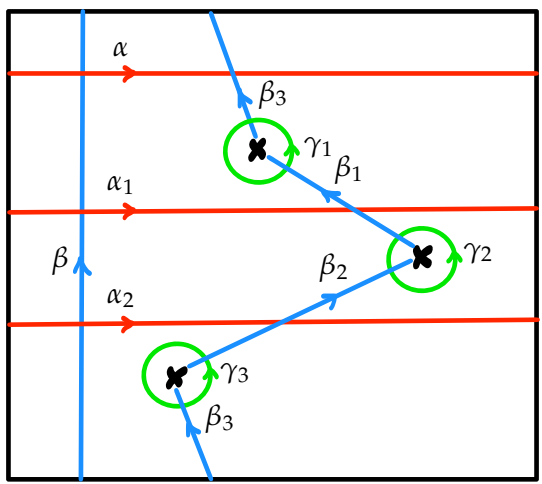

Following the analysis of~\cite{Amariti:2015dxa,Amariti:2016bxv}, we can study the charge lattices 
in terms of the \M2--lines
by introducing the notion of the fundamental group.
This is the  set of homotopy
classes of curves, where two closed curves are said to be 
homotopic if one can be continuously deformed into the other. 
A possible  presentation of the fundamental group of the \(r\)--punctured torus  \(\pi_1(\Sigma_{1,r})\) is obtained in terms of the \(\alpha \) and \(\beta\) cycles of the torus, plus a set of \(r\) cycles \(\set{\gamma_l}_{l=1}^r\) that go around each puncture \(P_l\) (see Figure~\ref{fig:cycles-and-paths}), together with the condition that there is a non-contractible line of trivial homology that can be written either as the commutator of \(\alpha\) and \(\beta\) or as the product of the \(\gamma_l\):
\begin{equation}
  \pi_1(\Sigma_{1,r}) = \braket{\alpha, \beta, \gamma_1, \gamma_2, \dots, \gamma_{r}}{\comm{\alpha}{\beta} = \gamma_1 \gamma_2 \dots \gamma_r} .
\end{equation}
This relation can be used to rewrite \(\gamma_r\) as a function of the other generators:
\begin{equation}
  \gamma_r = (\gamma_1 \dots \gamma_{r-1})^{-1} \comm{\alpha}{\beta} ,
\end{equation}
so that \(\pi_1(\Sigma_{1,r})\) is the free group of \(r+1\) generators,
\begin{equation}
  \pi_1(\Sigma_{1,r}) = \freebraket{\alpha,\beta,\gamma_1, \dots \gamma_{r-1}} ,
\end{equation}
endowed with the symplectic structure \(\intersection*{\cdot}{\cdot}\) describing the intersection of two curves, which in this basis reads:
\begin{align}
  \intersection*{\alpha}{\beta} &= 1, & \intersection*{\alpha}{\gamma_l} &= 0, & \intersection*{\beta}{\gamma_l} &=0, & \intersection*{\gamma_l}{\gamma_{l'}} &= 0. 
\end{align}

There is an alternative basis for the free group which is convenient for our problem.
Consider a set of \(r\) \(\alpha\)-cycles \(\alpha_l\) defined as (see Figure~\ref{fig:cycles-and-paths})
\begin{equation}
  \begin{cases}
    \alpha_l = \alpha \gamma_1 \dots \gamma_l & \text{for \(l = 1, \dots, r -1\)}, \\
    \alpha_r = \alpha.
  \end{cases}
\end{equation}
We can invert the relation and write
\begin{equation}
  \gamma_l = \alpha_{l-1}^{-1} \alpha_{l}
\end{equation}
to show that the fundamental group can be recast in the form
\begin{equation}
  \pi_1(\Sigma_{1,r}) = \freebraket{\alpha_1, \dots, \alpha_r, \beta} ,
\end{equation}
with the symplectic structure
\begin{align}
  \intersection*{\alpha_l}{\beta} &= 1, & \intersection*{\alpha_l}{\alpha_{l'}} &= 0.
\end{align}
% \begin{figure}
%   \centering
%   \begin{tikzpicture}
%     \begin{scope}[shift={(0,0)}]
%       \node at (0,0) {\includegraphics[width=.3\linewidth]{3pt-torus-basis-1-2d}};
%       \node at (-2.4,1.9) {\(\alpha\)};
%       \node at (-.7,2.3) {\(\beta\)};
%       \node at (-1.5,-.3) {\(\gamma_2\)};
%       \node at (-1.5,-1.7) {\(\gamma_1\)}; 
%       \node at (0,-2.5){(a.1)};
%     \end{scope}
%     \begin{scope}[shift={(.5\linewidth,0)}]
%       \node at (0,0) {\includegraphics[width=.35\linewidth]{3pt-torus-basis-1-3d}};
%       \node at (0.2,1.8) {\(\beta\)};
%       \node at (2.4,1.1) {\(\gamma_1\)};
%       \node at (2,-1.5) {\(\alpha\)};
%       \node at (-1.5,1.9) {\(\gamma_2\)};
%       \node at (0,-2.5){(a.2)};
%     \end{scope}
%     \begin{scope}[shift={(0,-6cm)}]
%       \node at (0,0) {\includegraphics[width=.3\linewidth]{3pt-torus-basis-2-2d}};
%       \node at (-2.4,1.9) {\(\alpha\)};
%       \node at (-.7,2.3) {\(\beta\)};
%       \node at (-2.4,.5) {\(\alpha_2\)};
%       \node at (-2.4,-1) {\(\alpha_1\)};
%       \node at (0,-2.5){(b.1)};
%     \end{scope}
%     \begin{scope}[shift={(.5\linewidth,-6cm)}]
%       \node at (0,0) {\includegraphics[width=.35\linewidth]{3pt-torus-basis-2-3d}};
%       \node at (-.1,1.8) {\(\beta\)};
%       \node at (2.2,-1.5) {\(\alpha\)};
%       \node at (.6,2.2) {\(\alpha_1\)};
%       \node at (-2.6,-.6) {\(\alpha_2\)};
%       \node at (0,-2.5){(b.2)};
%     \end{scope}    
%   \end{tikzpicture}
%   \caption{Cycles generating a basis for the fundamental group of the torus with \(r\) punctures.}
%   \label{fig:pi-basis}
% \end{figure}
The homology of a curve \(\mathcal{C}\) can be expressed in terms of either basis as
\begin{equation}
  [\mathcal{C}] = m [\beta] + e [\alpha] + \sum_{l=1}^{r-1} \lambda_l [\gamma_l] = m [\beta] + \sum_{l=1}^r e_l [\alpha_l],
\end{equation}
which provides the map between the coefficients:
\begin{equation}
  \begin{cases}
    e_1 = \lambda_1 - \lambda_2, \\
    e_2 = \lambda_2 - \lambda_3, \\
    \hspace{1em}\vdots \\
    e_{r-2} = \lambda_{r-2} - \lambda_{r-1}, \\
    e_{r-1} = \lambda_{r-1}, \\
    e_r = e - \lambda_1.
  \end{cases} % &&
              %    \begin{cases}
              %      e = e_1 + e_2 + \dots + e_r \\
              %      \lambda_1 = e_1 + e_2 + \dots e_{r - 1} + \mu \\
              %      \lambda_2 = e_2 + e_3 + \dots e_{r - 1} + \mu\\
              %      \hspace{1em}\vdots \\
              %      \lambda_r = \mu
              %    \end{cases}
\end{equation}
% ❌ where \(\mu \) is a free constant, corresponding to the fact that the cycle \(\gamma_r\) is not independent of the others.
The intersection number of two curves \(\mathcal{C}\) and \(\mathcal{C}'\) is then
\begin{equation}
  \intersection*{\mathcal{C}}{\mathcal{C}'} = \pqty{\sum_{l=1}^r e_l} m' - \pqty{ \sum_{l=1}^r e_l'} m  = e m' - e' m .
\end{equation}
This reproduces precisely the structure of the charges in the gauge theory. Since there is only one \(\beta\)-cycle\footnote{The asymmetry between \(\alpha\)-cycles and \(\beta\)-cycles is related to the \tIIA version of the geometry, where the \(\alpha\)-cycles become non-geometric and the punctured torus reduces to the necklace quiver.}, there is only one magnetic charge. The \(r\) \(\alpha\)-cycles correspond to the \(r\) electric charges and the \ac{dsz} condition is the intersection number between two geodesics on the Riemann surface which only depend on how many times the curve wraps the \(\alpha\) and the \(\beta\) cycles, \emph{i.e.} the sum of the electric charges and the unique magnetic charge.

\bigskip

Now that the geometric structure of the problem is set up, we have to consider
the multiple cover of the M--theory 
torus by the \M5--brane to reproduce the stack of $N$ \D4--branes in the \tIIA description and ultimately the non-Abelian 
$SU(N)_l$ gauge factors on the field theory side.
By studying the intersection of the cycles introduced above in the covering geometry and their projection 
to the field theoretical charges, we will be able to construct the lattices via the geometric 
analysis.

An \(N\)-cover of the \(r\)-punctured torus \(\Sigma^N_{1,r}\) is a torus with \(N \times r \) punctures (Riemann--Hurwitz). A given cover is identified by its fundamental group, which is a subgroup of index \(N\) of \(\pi_1(\Sigma_{1,r})\). These subgroups are classified in terms of maps from \(\pi_1(\Sigma_{1,r})\) to the symmetric group of \(N\) elements \(\mathfrak{S}_N\) and can be always put into the form
\begin{equation}
  \pi_1(\Sigma^N_{1,r}) = \braket{\alpha^k, \alpha^i \beta^{k'}, \gamma_{1,1}, \dots, \gamma_{1,N}, \gamma_{2,1}, \dots, \gamma_{2,N}, \gamma_{r,1}, \dots \gamma_{r,N}}{\comm{\alpha^k}{\alpha^i \beta^{k'}} = \prod_{l=1}^r\prod_{p=1}^N \gamma_{l,p}},
\end{equation}
where \(\gamma_{l,p}\) can be written as:
\begin{equation}
  \gamma_{l,p} = \Ad_{\lambda_{l,p}} \gamma_l = \lambda_{l,p} \gamma_l \lambda_{l,p}^{-1} 
\end{equation}
for some \(\lambda_{l,p} \in \pi_1(\Sigma_{1,r})\), chosen such that the relation in the presentation of the fundamental group of the cover is equivalent to the relation in the fundamental group of the base:
\begin{equation}
  [\alpha^k, \alpha^i \beta^k] ( \prod_{l=1}^r\prod_{p=1}^N \gamma_{l,p})^{-1} = [\alpha, \beta] ( \prod_{l=1}^r\gamma_{l})^{-1} = 1 .
\end{equation}
The integers \(k,k',i\) satisfy the relations
\begin{equation}
  \begin{cases}
    k k' = N \\ 
    0 \le i < k .
  \end{cases}
\end{equation} 
For fixed \(N\) there are \(\sigma_1(N)\) such covers, where \(\sigma_1\) is the divisor function, \emph{i.e.} the sum over all the divisors of \(N\): \(\sigma_1(N) = \sum_{d | N} d\). This is to be compared with the results of the previous section: once more we see that the geometric structure precisely reproduces the results of the gauge theory.

The cover \(\Sigma^N_{1,r}\) inherits a symplectic form from the base, given by
\begin{align}
  \intersection*{\alpha^k}{\alpha^i\beta^{k'}} &= N, & \intersection*{\alpha^k}{\gamma_{l,p}} &= 0, & \intersection*{\alpha^i \beta^{k'}}{\gamma_{l,p}} &=0, & \intersection*{\gamma_{l,p}}{\gamma_{l',p'}} &= 0. 
\end{align}
This means that if we take two closed curves \(\mathcal{C}^N\) and \({\mathcal{C}^N}'\) on \(\Sigma^N_{1,r}\), their symplectic product, counting how many times the projections of the curves will intersect on the base \(\Sigma_{1,r}\) is given by
\begin{equation}
  \label{interesction}
  \intersection*{\mathcal{C}^N}{{\mathcal{C}^N}'} = N \pqty{\pqty{\sum_{l=1}^r e_l} m' - \pqty{ \sum_{l=1}^r e_l'} m} .
\end{equation}
This fully reproduces the \ac{dsz} condition of Eq.~\eqref{DSZ-with-hypers}.

\bigskip

We have studied the homologies of the closed curves in the multiple covering space and interpreted these curves as bound states of 
\ac{Wline}s and \ac{Hline}s on the field theory side.
The situation is analogous to the one discussed in~\cite{Amariti:2015dxa}.
Again, the intersection number of these curves becomes the \ac{dsz} condition on the field theory side.
Note an interesting aspect of this quantization condition derived from M--theory: on the field theory side, 
the \ac{dsz} condition for the product of gauge groups in~Eq.\eqref{DSZ-general-An} is different from the one derived coming from the intersection of the lines in~Eq.\eqref{interesction}.
They become the same if we consider the presence of the hypermultiplets, because 
this fixes $m_i=m$ in~Eq.(\ref{DSZ-general-An}).
This is expected because the presence of the punctures in the geometry 
translates into the presence of the hypermultiplets in the field theory description.

This concludes our discussion of the derivation of the lattices for the elliptic models
from the M--theory description. 
We have shown how to interpret the charges of the lines in the geometric language and that
the study of the intersection numbers of closed curves in the geometry reproduces the 
field theory constraints imposed by the mutual locality condition.

%%%%%%%%%%%%%%%%
\section{S--duality}
\label{duality}

In this section, we discuss the structure of the S--duality group and its action on the lattices.

Let us start by discussing  the situation without punctures. This corresponds to the usual $\mathcal{N}=4$
\ac{sym} theory and the S--duality group corresponds to the action of the modular group $SL(2,\setZ)$ 
on the complex structure of the torus, $\tau$. The generators of this group act as
$S: \tau \rightarrow -1/\tau$ and $T: \tau \rightarrow \tau+1$. The action of these generators on a dyon with charge $(e,m)$ 
is 
\begin{align}
S:(e,m) &\rightarrow (-m,e),\\
T:(e,m) &\rightarrow (e+m,m).
\end{align}
This action corresponds to an exact duality on the string coupling.

When we add the punctures, there still is an \(SL(2, \setZ)\) acting on the string coupling but now we have \(r\) components, each with its own gauge group that a priori has an \(SL(2, \setZ)\) symmetry. This means that the full S--duality group must be more intricate. %
% Next, we consider the presence of the orbifold: there is still an $SL(2,\setZ)$ acting on the string coupling
% but the compactification induces $r$ components, each associated to a different gauge theory. 
% Despite the existence of a single $SL(2,\setZ)$ symmetry in the string theory picture, 
% the full S--duality group is more intricate here. 
This situation is clarified by the M--theory description.

\bigskip

We have seen that the four-dimensional gauge theory can be regarded as a reduction from six dimensions on the multiple cover of a punctured torus \(\Sigma_{1,r}\). 
If we apply an isomorphism of the torus before the reduction, this will in general lead to a different four-dimensional gauge theory that is related to 
the previous one by S--duality. In other words, the action of the ``symmetries'' of the Riemann surface (the mapping class group \(\Mod(\Sigma_{1,r})\)) 
will produce all the possible phases of a given necklace quiver gauge theory.

The mapping class group of a punctured Riemann surface
\(\Sigma_{g,r}\) is decomposed into the product of the pure mapping
class group \(\PMod(\Sigma_{g,r})\) that leaves each puncture
invariant and the permutation group \(\mathfrak{S}_r\) acting on the punctures~\cite{Benson:2011}. More precisely, the following is a short exact sequence:
\begin{equation}
  1 \to \PMod(\Sigma_{g,r}) \to \Mod(\Sigma_{g,r}) \to \mathfrak{S}_r \to 1 .
\end{equation}
It follows that a generating set for \(\Mod(\Sigma_{g,r})\) is given by a generating set for \(\PMod(\Sigma_{g,r})\) together with a set of elements in \(\Mod(\Sigma_{g,r})\) that project to generators of \(\mathfrak{S}_r\), \emph{i.e.} the \(r-1\) transpositions of two consecutive punctures.

\begin{figure}
  \centering
  \begin{tikzpicture}
    \node at (0,0)  {\includegraphics[width=.9\textwidth]{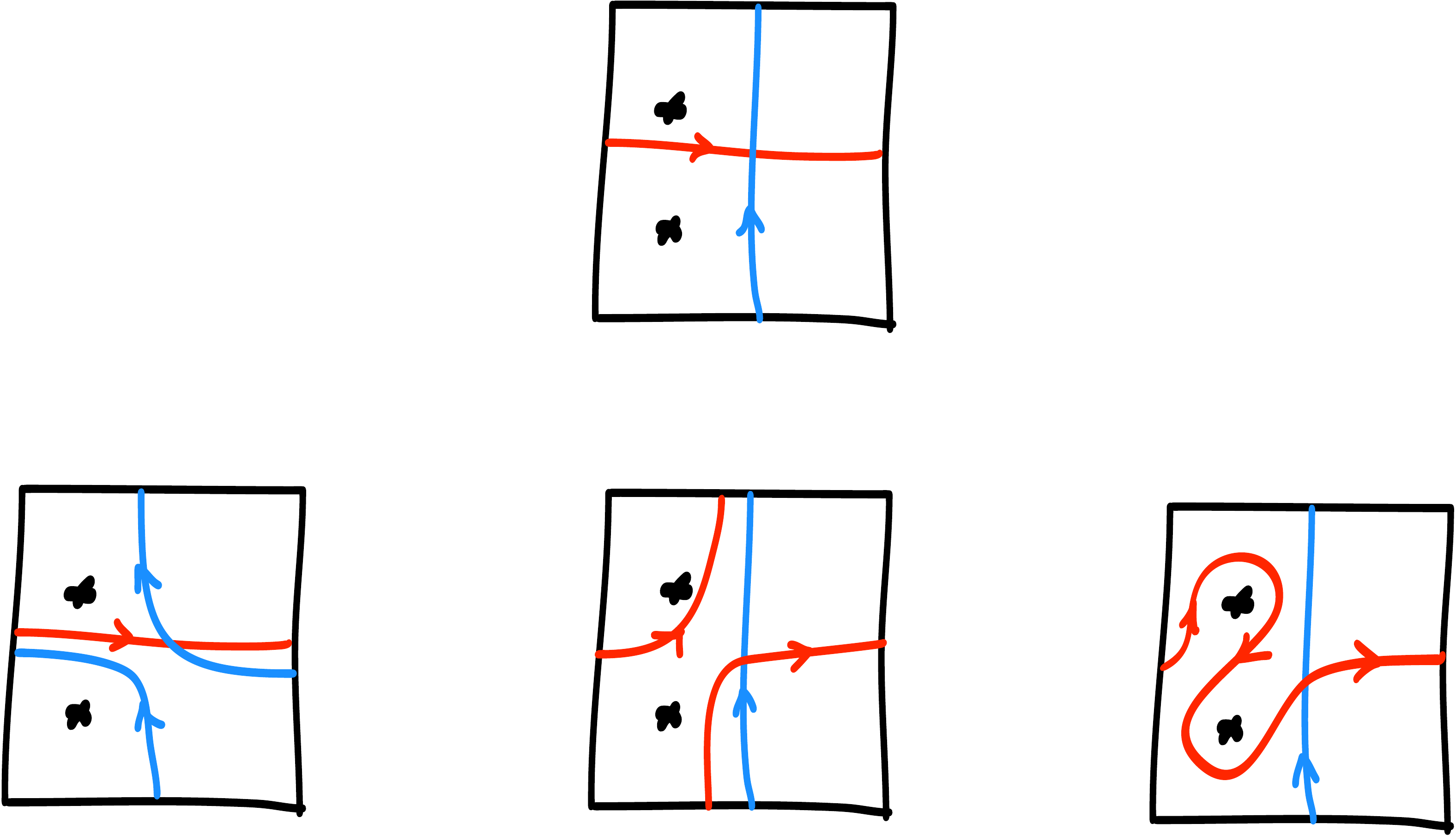}};
    \node at (-1.4,2.4) {\(\alpha_l\)};
    \node at (0.2,3.9) {\(\beta\)};
    \node at (-.5,3.1) {\(P_l\)};
    \node at (-.6,1.2) {\(P_{l+1}\)};

    \node at (-5.3,-4) {(a)};
    \node at (0,-4) {(b)};
    \node at (5,-4) {(c)};
  \end{tikzpicture}
  \caption{Action of the mapping class group generators in terms
    of Dehn twists. One way of performing the twist consists in resolving the intersection of the two curves so that the twisted curve is surgered and the surgered arc turns left at the intersection. In 
 (a) we see the twist  \(T_l\) acting on \(\beta\), in (b) the
    twist  \(T_\beta\) acting on \(\alpha_l\). Figure (c) represents the half-twist of \(\alpha_l\) around a curve
    that encloses the punctures \(P_l\) and \(P_{l+1}\), giving the permutation \(\sigma_l\).}
  \label{fig:Dehn-generators}
\end{figure}
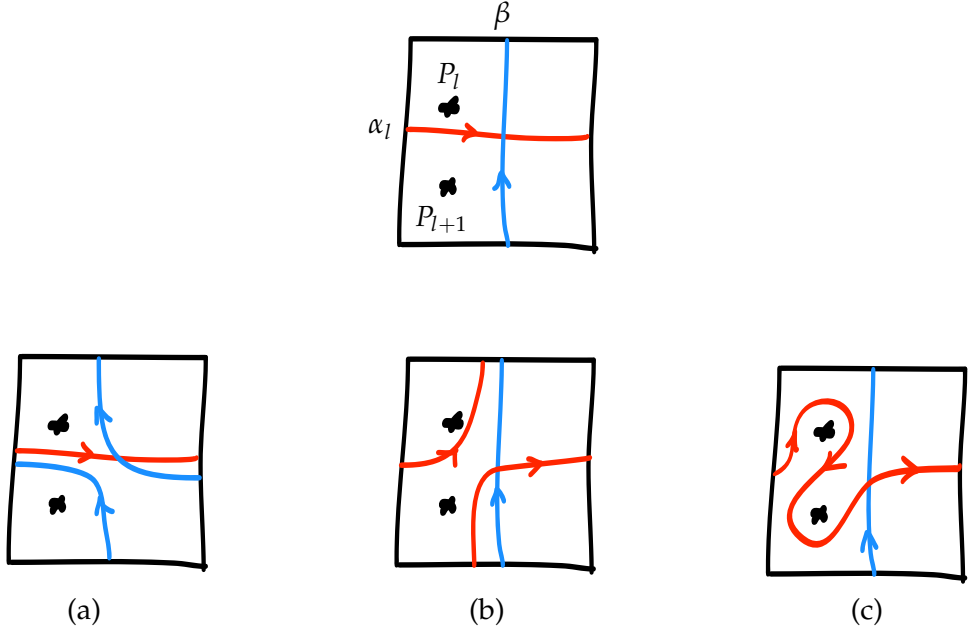

The group \(\PMod(\Sigma_{g,r})\) is generated by a set of Dehn twists which, for the punctured torus \(\Sigma_{1,r}\), are around the cycles \(\alpha_l\) and \(\beta\) (see Figure~\ref{fig:Dehn-generators}(a) and~\ref{fig:Dehn-generators}(b)). They act on the generators of \(\pi_1(\Sigma_{1,g})\) as follows:
\begin{align}
  T_n &: \set{\alpha_l, \beta} \mapsto \set{\alpha_l, \beta  \alpha_n^{-1}  }, & \text{ for \(n = 1, \dots, r\)} \\
  T_\beta &: \set{\alpha_l, \beta} \mapsto \set{\alpha_l \beta, \beta }.
\end{align}
The transposition of two punctures, say \(P_n\) and \(P_{n+1}\),
corresponds to a Dehn half-twist around a curve that encloses the two punctures (see Figure~\ref{fig:Dehn-generators}(c)) and acts on the fundamental group as follows:
\begin{equation}
  \sigma_n : \set{\alpha_l, \beta} \mapsto \set{\alpha_1, \dots, \alpha_{n-1}, \alpha_{n-1} \alpha_n^{-1} \alpha_{n+1}, \alpha_{n+1}, \dots, \beta}.
\end{equation}

A minimal set of generators for \(\Mod(\Sigma_{1,r})\) is given by two elements from \(\PMod(\Sigma_{1,r})\), \(T = T_r \) and \(S = \pqty{T_\beta T_r T_\beta}^{-1}\), together with two from \(\mathfrak{S}_r\) acting as follows:
\begin{align}
  T &: \set{\alpha_l, \beta} \mapsto \set{\alpha_1, \dots, \alpha_r, \beta \alpha_r^{-1}},\\
  S &: \set{\alpha_l, \beta} \mapsto \set{\alpha_r^{-1} \alpha_1 \beta^{-1}, \alpha_r^{-1} \alpha_2 \beta^{-1}, \dots, \alpha_r^{-1} \alpha_{r-1} \beta^{-1} , \beta^{-1}, \alpha_r} ,\\
  \sigma_1 &: \set{\alpha_l, \beta} \mapsto \set{\alpha_r \alpha_1^{-1} \alpha_2, \alpha_2, \dots, \alpha_r, \beta}, \\
  \omega &: \set{\alpha_l, \beta} \mapsto \set{\alpha_2, \alpha_3, \dots, \alpha_r, \alpha_1, \beta} .
\end{align}
Observe that \(\omega\) cyclically permutes all the punctures, it is the generator of the cyclic group \(\setZ_r = \braket{\omega}{\omega^r = 1}\). This is the symmetry group of the affine \(\hat A_{r - 1}\) Dynkin diagram, which has the same shape as our necklace quiver. In this sense, we can think of \(\setZ_r\) as of a classical symmetry (realized geometrically in \tIIA), which is enhanced by quantum effects to \(\Mod(\Sigma_{1,r})\) (realized geometrically in M--theory).

Each closed curve on the cover \(\Sigma^N_{1,r}\) corresponds to a \ac{bps} line operator in the necklace quiver gauge theory, whose  central charge is\footnote{These are not the integrals used to define the metric on the moduli space of the theory. See Appendix~\ref{sec:moduli-space-metric} for a discussion.}
\begin{equation}
  \label{eq:central-charge}
  Z = \sum_{l = 1}^r e_l  a^l + m  a_D ,
\end{equation}
where \( a^l\) and \( a_D\) are the integrals of the Seiberg--Witten
differential \(\lambda\) around the cycles \(\alpha_l\) and \(\beta\):
\begin{align}
   a^l &= \int_{\alpha_l} \lambda ,&  a_D &= \int_{\beta} \lambda .
\end{align}
An element \(M \in \Mod(\Sigma_{1,r})\) acts as a matrix on the vector
\(( a^1, \dots,  a^r,  a_D)\):
\begin{equation}
  M \in \Mod(\Sigma_{1,r}) : \mqty( a^1 \\ \vdots \\  a^r \\  a_D) \mapsto M \mqty( a^1 \\ \vdots \\  a^r \\  a_D) = \mqty( {a^1}' \\ \vdots \\  {a^r}' \\   a_D' ) .
\end{equation}
The elements of the mapping class group are invertible. So there exists a matrix \(W = M^{-1}\) that, acting on the charge vector \((e_1, \dots, e_r, m)\) on the right, preserves the central charge:
\begin{equation}
  Z = (e_1, \dots, e_r, m)  \mqty( a^1 \\ \vdots \\  a^r \\  a_D) = (e_1, \dots, e_r, m) W M \mqty( a^1 \\ \vdots \\  a^r \\  a_D) = (e_1', \dots, e_r', m')  \mqty( {a^1}' \\ \vdots\\  {a^r}' \\  a_D') . 
\end{equation}
We have found a symmetry of the \emph{full theory} under which a \ac{bps} state of charge \((e_1, \dots, e_r, m)\) is mapped to another state with charge \((e_1', \dots, e_r', m')\) when the \((a^l, a_D)\) are mapped to \({a^l}', a_D'\). For the generators of \(\Mod(\Sigma_{1,r})\) we find explicitly
\begin{equation}
  \begin{aligned}
    S&: (e_1, \dots, e_r; m) \mapsto (e_1, \dots, e_{r-1}, e_r - e - m; e), \\
    T&: (e_1, \dots, e_r; m)  \mapsto (e_1, \dots, e_{r-1}, e_r + m; m), \\
    % ❌ T_l&: (e_1, \dots, e_r; m)  \mapsto (e_1, \dots, e_{l-1} - m, e_l + m, \dots ; m) \\
    % ❌ t_l&: (e_1, \dots, e_r; m)  \mapsto (e_1, \dots, e_{l-1} - e, e_l + e, \dots ; m) \\
    \sigma_1&: (e_1, \dots, e_r; m)  \mapsto (-e_1, e_1 + e_2, e_3 \dots, e_{r-1}, e_1 + e_r; m), \\
    \omega &: (e_1, \dots, e_r; m) \mapsto (e_2, e_3, \dots, e_r, e_1; m),
  \end{aligned}
\end{equation}
% ❌ The mapping class group \(\Mod(\Sigma_{1,r})\) is generated by the transformations \(\set{S,T,T_l,t_l,s_l}\) that act on the homology of a curve \(\mathcal{C}\) in the \(\freebraket{a_l,b}\) basis as follows~\cite{Birman1,Birman2}:
where \(e = e_1 + \dots +e_r\) is the total electric charge. % Note that the transformations satisfy relations such that  \(\set{S,T,s_1,\omega}\) is a minimal set of generators.

\bigskip 
Now we can give a physical interpretation for the action of the mapping class group.
% ❌ This basis has been chosen to give an immediate physical interpretation of the transformations:
\begin{itemize}
\item The operators \(S \) and \(T\) act like \(SL(2,\setZ)\) transformations on the \emph{total electric charge} and on the magnetic charge:
  \begin{equation}
    \begin{aligned}
      S &: (e;m) \mapsto (-m; e), \\
      T &: (e;m) \mapsto (e + m; e).
    \end{aligned}
  \end{equation}
  
  Since a phase is identified by the allowed values of \(e\) and \(m\), these operators do in general map one phase to another.\\
   Observe that they \emph{do not} satisfy the usual \(SL(2, \setZ)\) relations, though. In fact we find that
  \begin{equation}
    S^2 = \pqty{ST}^3 : (e_1 \dots, e_r; m) \mapsto (e_1, \dots e_{r-1}, -2 e + e_r; -m),
  \end{equation}
  so that
  \begin{equation}
    S^4 = \pqty{ST}^6 = \Id.
  \end{equation}
  These transformations generate the \(SL(2,\setZ)\) discussed above: it is \emph{independent} of the number of punctures. 
  The physical interpretation of this  \(SL(2, \setZ)\) is clarified by the geometric description: in principle, one could define an 
   \(SL(2,\setZ)\)  for each gauge group and imagine the notion 
   of the ``diagonal'' \(SL(2,\setZ)\) (see~\cite{Halmagyi:2004ju} for a similar discussion). Here we see that this
 is \emph{not the correct picture}. The  \(SL(2,\setZ)\) subgroup of the mapping class group does indeed select one of the groups, \emph{i.e.} it acts only on one
 of the $\alpha_l$ cycles and on the cycle $\beta$. 
 The $r$ different choices of the gauge group are related by the action of $\omega$.
  \item The operators \(\sigma_1\) and \(\omega\) do not change \(e \) or \(m\) but change 
  the distribution of the electric charge among the gauge groups. These transformations map a state in a given phase into another state in the same phase. This corresponds to the intuition that a permutation of the punctures (the \NS5--branes) does not change the total number of \M2--branes that are reduced to fundamental strings, but only how the \F1s are distributed among the stacks of \D4--branes.
\end{itemize}

We can now completely describe the phases of a necklace quiver with algebra \(A_{N-1} \oplus \dots \oplus A_{N-1}\). Each phase is identified by a two-dimensional lattice with components \(e\) and \(m\) corresponding to the total electric charge \(e = e_1 + \dots + e_r\) and the magnetic charge \(m\) of the allowed \ac{WH} lines. The operators \(\set{\sigma_1, \omega}\) leave the lattice invariant, while \(S\) and \(T\) map in general one lattice into another. This leads precisely to the same phase space as the one of the \(A_{N-1}\) \(\mathcal{N} = 4\) gauge theory~\cite{Aharony:2013hda,Amariti:2015dxa}. For fixed \(N\), there are \(\sigma_1(N)\) (with $\sigma$ the divisor function) phases that are arranged into orbits of \(S\) and \(T\). The number of distinct orbits is given by the number of ways in which \(N\) can be written in the form \(N = n_1 \times n_2^2\) in terms of two integers \(n_1 \) and \(n_2\)~\cite{Amariti:2015dxa}.

%%%%%%%%%%%%
\section{Example: the quiver \(A_1 \oplus A_1 \oplus A_1 \)}\label{sec:ex}

Consider the case \(N = 2\), \(r = 3\) of a necklace quiver with algebra \(A_1 \oplus A_1 \oplus A_1\). A generic \ac{WH} line has charges \((e_1, e_2, e_3 ; m) \in (\setZ_2)^4\). Two such lines can coexist in the same phase (\emph{i.e.} the same gauge theory with fixed gauge group) if
\begin{equation}
  \pqty{e_1 + e_2 + e_3} m' - \pqty{e_1' + e_2' + e_3'} m = e m' - e' m = 0 \mod 2.
\end{equation}
We have three distinct possibilities, corresponding to the three lattices with charges \((e;m)\) generated by \(\Gamma_{2;1,0} = \freebraket{(1,0),(0,2)}\), \(\Gamma_{2;2,0} = \freebraket{(2,0),(0,1)}\) and \(\Gamma_{2;2,1} = \freebraket{(2,0),(1,2)}\). These are the homologies of the closed curves living on the three double covers of the Riemann surface \(\Sigma_{1,3}\). For example, the lattice \(\Gamma_{2;2,0}\) describes the homologies in the cover \(\Sigma^2_{1,3}\) with fundamental group
\begin{equation}
  \Sigma^2_{1,3} = \braket{\alpha^2, \beta, \Ad_\alpha \gamma_1, \Ad_\alpha \gamma_2, \Ad_\alpha \gamma_3, \gamma_1, \gamma_2, \gamma_3}{\comm{\alpha}{\beta} = \gamma_1 \gamma_2 \gamma_3} .
\end{equation}
In fact, the projection on \(\Sigma_{1,3}\) of a closed curve \(\mathcal{C}^2\) in \(\Sigma^{2}_{1,3}\) has homology
\begin{equation}
  [\mathcal{C}^2] = 2p [\alpha] + q [\beta] + \lambda_1 [\gamma_1] + \lambda_2 [\gamma_2] + \lambda_3 [\gamma_3],
\end{equation}
corresponding to a \ac{WH} line of charge \((\lambda_1 - \lambda_2, \lambda_2 - \lambda_3, 2p + \lambda_3 - \lambda_1 ; q)\).

The transformations \(\set{\sigma_1, \omega}\) act on this state as
\begin{align}
  \sigma_1 &: (e_1, e_2, 2 p - e_1 - e_2 ; q) \mapsto (-e_1, e_1 + e_2, 2p - e_2 ; q), \\
  \omega &: ( e_1, e_2, 2 p - e_1 - e_2 ; q) \mapsto (e_2, 2p - e_1 - e_2, e_1 ; q),
\end{align}
and are endomorphisms of the lattice.

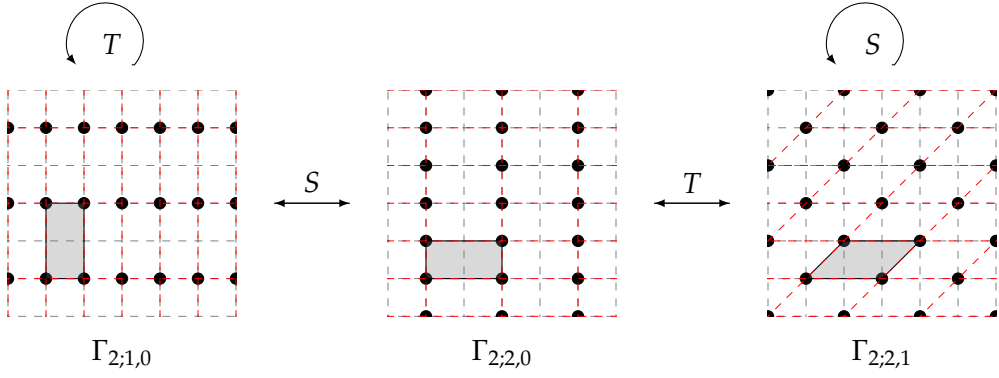
\begin{figure}
  \centering
  \begin{tikzpicture}
    \node at (-5,0) {\lattice{1}{0}{2}};
    \node at (0,0) {\lattice{2}{0}{1}};
    \node at (5,0) {\lattice{2}{1}{1}};
    \node at (-5,-2) {\(\Gamma_{2;1,0}\)};
    \node at (0,-2) {\(\Gamma_{2;2,0}\)};
    \node at (5,-2) {\(\Gamma_{2;2,1}\)};
    \draw[-latex] (-2,0) -- node[above] {\(S\)} ++ (-1,0);
    \draw[-latex] (-3,0) -- (-2,0);
    \draw[-latex] (2,0) -- node[above] {\(T\)} ++ (1,0);
    \draw[-latex] (3,0) -- (2,0);

    \draw[-latex, join=round] (-4.5,1.5) ++(138:5mm) node[above left]{\(T\)} --++(180:-1pt) arc (-40:220:5mm);
    \draw[-latex, join=round] (5.5,1.5) ++(138:5mm) node[above left]{\(S\)} --++(180:-1pt) arc (-40:220:5mm);

  \end{tikzpicture}
  \caption{The three charge lattices (and phases) of the \(A_1 \oplus A_1 \oplus A_1\) necklace quiver. The three phases are in the same orbit of the mapping class group \(\Mod(\Sigma_{1,3})\).}
  \label{fig:A1-A1-A1-lattices}
\end{figure}

The transformations \(S \) and \(T\) map the lattice
\(\Gamma_{2;2,0}\) respectively to the lattices \(\Gamma_{2;1,0}\) and
\(\Gamma_{2;2,1}\), showing that the three phases belong to the same
S--duality orbit:
\begin{align}
  S &: (e_1, e_2, 2 p - e_1 - e_2 ; q) \mapsto (e_1, e_2, -q - e_1, -e_2; 2p) \in \Gamma_{2;1,0}, \\ 
  T &: (e_1, e_2, 2 p - e_1 - e_2 ; q) \mapsto (e_1, e_2, 2 p + q - e_1 - e_2 ; q) \in \Gamma_{2;2,1}.
\end{align}
See Figure~\ref{fig:A1-A1-A1-lattices} for the full diagram showing the complete action of \(\PMod(\Sigma_{1,3})\) on the three lattices of the \(A_1 \oplus A_1 \oplus A_1\) necklace quiver.

%%%%%%%%%%%%%%%%%%%%%
\section{Further directions}
\label{conclusions}

In this paper we have studied the global properties of $\mathcal{N}=2$ necklace quiver gauge theories
with $r$ nodes. They can be understood in terms of the charge lattices of mutually local bound states of \acp{Wline} and \acp{Hline}.
 We find that they can be formulated as two-dimensional lattices which correspond 
 to the ones obtained in $\mathcal{N}=4$ \ac{sym}. 
We have reinterpreted the analysis in a geometric language by studying the uplift of this system to M--theory.
In this picture, the problem reduces to studying the homologies of closed \M2--lines on the \(N\)-cover of
a torus with $r$ punctures.
We have reproduced the field theory results by introducing the notion of the fundamental group.
Finally, we have shown how to connect different lattices by S--duality, corresponding to the action 
of the mapping class group of the Riemann surface. 
The latter is decomposed % in the field theory language 
into the combined action of the $SL(2,\setZ)$ symmetry 
on the torus and of the permutation and shift symmetries of the punctures.
% ❌ As a net effect 
Only the generators of $SL(2,\setZ)$ act non-trivially on the lattices, which can be organized into separate orbits 
% ❌ Another result consist of the existence of orbits in 
of the S--duality group, just as in $\mathcal{N}=4$ \ac{sym}.

\bigskip

Our geometrical analysis can also be useful for class $S$ theories. These theories are constructed by gluing fundamental $\mathcal{N}=2$
$T_N$ blocks, with $SU(N)^3$ global symmetry.
In the M--theory description, these blocks represent spheres with three punctures and the gluing operation corresponds to 
the gauging of the global symmetries. The four-dimensional theories are in general non-Lagrangian and are obtained 
by a partially twisted compactification of the Riemann surface obtained by gluing $T_N$ blocks. 
The M--theory description has been used in~\cite{Amariti:2016bxv}
to derive the global properties of these four-dimensional gauge theories, but the analysis was restricted to the case
of compact Riemann surfaces. Here, we have considered the presence of punctures in similar geometries. 
It would be interesting to generalize our current understanding  to the case of class 
S theories with generic punctures.

Another interesting line of research consists in studying $\mathcal{N}=1$ theories.
One can indeed generalize the analysis to $\mathcal{N}=1$ theories with an M--theory origin.
It can be done by giving some masses to the adjoints in the elliptic models (\emph{e.g.} by embedding the construction in a fluxtrap background~\cite{Hellerman:2011mv,Hellerman:2012zf,Orlando:2013yea}), adding fluxes (see \emph{e.g.}~\cite{Lambert:2013lxa,Lambert:2014fma}) or by looking at some generalizations
of the class S theories, like the Sicilian theories~\cite{Benini:2009mz} and the class $S_k$ theories~\cite{Gaiotto:2015usa,}.
In these cases, the possible lattices have to coincide with the ones studied here. This can be verified by reproducing our $\mathcal{N}=1$ field theory 
analysis of section~\ref{field}. As already observed there, this result is also expected from the brane description: there is a $U(1)$ 
symmetry, namely the center of mass of the stack of branes on which the gauge theory lives,  that 
decouples in the \ac{ir}. 
The different consistent factorizations of this $U(1)$ symmetry correspond to the various theories associated to the
same algebra~\cite{Moore:2014gua}.
% ❌ It may however be interesting to reproduce our M--theory analysis also for these geometries.

 Another extension of our discussion regards the classification of the lattices for theories with real gauge groups,
 corresponding to the presence of orientifold fixed points in the M--theory picture. This requires taking into account the effect of these fixed points in the fundamental group.
In the $\mathcal{N}=1$ case this analysis may have interesting consequences on the structure of the S--duality group.

\section*{Acknowledgments}
We are grateful to Philip Argyres and Mario Martone for discussions.
A.A. thanks the University of Cincinnati for hospitality during the final stages of this work. 
The work of S.R. is supported by the Swiss National Science Foundation
(\textsc{snf}) under grant number \textsc{pp}00\textsc{p}2\_157571/1.

\appendix
\section{The metric on the moduli space}
\label{sec:moduli-space-metric}

\(\mathcal{N} = 2\) supersymmetry fixes the metric on the moduli space to be of the form
\begin{equation}
  \dd{s}^2 = \Im[ \dd{  a_{D,l}} \dd{\overline{  a^l}}],
\end{equation}
where \( a^i\) and \( a_{D,i}\) are integrals of the Seiberg--Witten
one-form \(\lambda\) over some paths on the Riemann surface. In our
case of a torus with \(r\) punctures, they can be defined as
follows\footnote{See~\cite{Itoyama:1995uj} for an equivalent basis.}:
\begin{align}
   %a^r &= \int_{\alpha} \lambda , &  %a_{D,r} &= \int_\beta \lambda , &
   a^l &= \int_{\alpha_l} \lambda , &  a_{D,l} &= \int_{\beta_l} \lambda. %& \text{for \(l = 1, \dots, r-1\)},
\end{align}
\(\beta_l\) is the line that joins \(P_l\) to \(P_{l+1}\) with the
 convention \(P_{r+1} = P_1\) (see Figure~\ref{fig:cycles-and-paths}) and \(\alpha_r = \alpha\). These paths are chosen such that their non-vanishing intersections are
\begin{equation}
  % \intersection*{\alpha}{\beta_r} &= 1 , & 
                                           \intersection*{\alpha_l}{\beta_m} = \delta_{lm}.
                                       % \begin{cases}
                                       %   1 & \text{if \(l = m\)} \\
                                       %   0 & \text{otherwise.}
                                       % \end{cases}
\end{equation}
The action of the generators of \(\Mod(\Sigma_{1,r})\) on the integrals is
% Using the fact that \(\alpha_l = \alpha \gamma_1 \dots \gamma_l\) we can write the action of the generators of \(\PMod(\Sigma_{1,r})\) on the integrals:
\begin{align}
  &\begin{multlined}[.9\linewidth]
    T_{n} : ( a^1, \dots,  a^r,  a_{D,1}, \dots,  a_{D,r}) \mapsto \\ 
 ( a^1, \dots,  a^r,  a_{D,1}, \dots,  a_{D,n-1},  a_{D,n} -  a^n, a_{D,n+1} \dots, a_{D,r}),
  \end{multlined} \\
  &\begin{multlined}
    T_{\beta} : ( a^1, \dots,  a^r,  a_{D,1}, \dots,  a_{D,r}) \mapsto ( a^1 + a_{D}, \dots,  a^r +  a_{D},  a_{D,1}, \dots,  a_{D,r}) ,
  \end{multlined} \\
  &    \begin{multlined}[.9\linewidth]
      \sigma_n : ( a^1, \dots,  a^r,  a_{D,1}, \dots,  a_{D,r}) \mapsto \\
  (a^1, \dots, a^{n-1}, a^{n-1} - a^n + a^{n+1}, a^{n+1}, \dots, a^r, \\a_{D,1}, \dots, a_{D,n-2},
 a_{D,n-1} + a_{D,n}, - a_{D,n}, a_{D, n+1} +a_{D,n}, a_{D, n + 2} , \dots, a_{D,r}) ,
    \end{multlined}
\end{align}
where \(a_D = \sum_l a_{D,l}\).

% ❌ \begin{align}
% ❌   T_{n} :{}& ( a^1, \dots,  a^r,  a_{D,1}, \dots,  a_{D,r}) \mapsto \\ 
% ❌ &&  ( a^1, \dots,  a^r,  a_{D,1}, \dots,  a_{D,n-1},  a_{D,n} -  a^n, a_{D,n+1} \dots, a_{D,r})\nonumber \\
% ❌   T_{\beta} :{}& ( a^1, \dots,  a^r,  a_{D,1}, \dots,  a_{D,r}) \mapsto ( a^1 + a_{D,r}, \dots,  a^r +  a_{D,r},  a_{D,1}, \dots,  a_{D,r}) \\
% ❌   \sigma_n :{}& ( a^1, \dots,  a^r,  a_{D,1}, \dots,  a_{D,r}) \mapsto \\
% ❌ &  (a^1, \dots, a^{n-1}, a^{n-1} - a^n + a^{n+1}, a^{n+1}, \dots, a^r, a_{D,1}, \dots, a_{D,n-2},\nonumber\\
% ❌ & a_{D,n-1} + a_{D,n}, - a_{D,n}, a_{D, n+1} +a_{D,n}, a_{D, n + 2} , \dots, a_{D,r}).\nonumber
% ❌ \end{align}
%
% If we reorder the components in the vector as \(( a^r,  a^1, \dots,  a^{r-1},  a_{D,r},  a_{D,1}, \dots,  a_{D,r-1})\), we can write the \(T\) operators as block matrices:
% \begin{align}
%   T_n =
%   \begin{pmatrix}
%     1 & 0\\
%     0 & \ddots \\
%      \\
%     \\
%     & & & & & 0\\
%     & & & & 0 & 1 \\
%     -1 & \dots & -1 & 0 & \dots & 0 & 1 \\
%     \vdots & & \vdots \\
%     -1 & \dots & -1 & 0 & \dots & 0\\
%     0 \\
%     \vdots \\
%     0
%   \end{pmatrix}
% \end{align}
One can easily verify that the twists live in \(Sp(2r, \setZ)\), \emph{i.e.} they preserve the symplectic structure
\begin{equation}
  T^t \varepsilon T = \varepsilon,
\end{equation}
where \(\varepsilon\) is the matrix with components
\begin{equation}
  \varepsilon^i_{\phantom{i}j} =
  \begin{cases}
    1 & \text{if \(j = i + r\),} \\
    -1 & \text{if \( i = j + r\).}
  \end{cases}
\end{equation}
It follows that they leave the metric on the moduli space \(\dd{s}^2 = \Im[ \dd { a_{D,l}} \dd {\overline{ a^l}}]\) invariant.

\bigskip

The central charge of a \ac{bps} object in the theory can be written in terms of the \( a^l\) and \( a_{D,l}\) as
\begin{equation}
  Z =  e_l  a^l +  m^l  a_{D,l}.
\end{equation}
Taken separately, the integrals along the paths \(\beta_l\) diverge but the divergence coming from the puncture \(P_l\) appears with opposite signs in \(a_{D,l-1}\) and \(a_{D,l}\). This means that the central charge is finite if and only if all the coefficients \(m^l\) are equal. The result is that we can interpret the configuration in terms of \(m^l = m\) \M2--lines of finite length wrapping the cycle \(\beta\).
 % ❌ and  if we limit ourselves to \M2--lines of finite length we have necessarily \( m^l = 0\) for \(l = 1, \dots, r - 1\). 
In the \tIIA reduction this corresponds to having the same number of \D2--branes between each pair of \NS5s, \emph{i.e.} the magnetic charge of a \ac{bps} state must be the same for each of the gauge components. Once more we see a field-theoretical quantum condition resulting from a classical condition in M--theory. 
Since the only magnetic component remaining is \( m^l = m\) we can rewrite the central charge as in Eq.~\eqref{eq:central-charge} where \(a_D =  \sum_l a_{D, l}\).

\setstretch{.95}
\printbibliography

\end{document}